\documentclass[aps,pre,preprint,groupedaddress]{revtex4}
\usepackage{graphicx}
\usepackage{dcolumn}
\usepackage{bm}
\usepackage{algorithmic}
\usepackage{amsmath}
\usepackage{lineno}

\begin{document}


\title{Prescription-induced jump
distributions in multiplicative Poisson processes}


\author{Samir Suweis$^{1}$, Amilcare Porporato$^{2,1}$, Andrea Rinaldo$^{3,1}$, Amos Maritan$^{4}$}

\affiliation{$^1$ Laboratory of Ecohydrology ECHO/ISTE/ENAC,
Facult\'e ENAC,\'Ecole Polytechnique F\'ed\'erale Lausanne (EPFL)
Lausanne (CH) (samir.suweis@epfl.ch)
\\ $^2$ Department of Civil and Environmental Engineering Duke University Durham NC (USA)
(amilcare.porporato@epfl.ch) \\
$^3$ Dipartimento IMAGE Universit\`a di Padova I-35131 Padova
Italy (andrea.rinaldo@unipd.it)
\\
$^4$ Dipartimento di Fisica G. Galilei, INFN and CNISM Universit\`a di
Padova, Via Marzolo 8  I-35151 Padova  Italy
(maritan@pd.infn.it)\\}


\date{\today}

\begin{abstract}
Generalized Langevin equations (GLE) with multiplicative white
Poisson noise pose the usual prescription dilemma leading to
different evolution equations (master equations) for the probability
distribution. Contrary to the case of multiplicative gaussian
white noise, the Stratonovich prescription does not correspond to
the well known mid-point (or any other intermediate) prescription.
By introducing an inertial term in the GLE we show that the Ito and
Stratonovich prescriptions naturally arise depending on two time
scales, the one induced by the inertial term and the other
determined by the jump event. We also show that when the
multiplicative noise is linear in the random variable one
prescription can be made equivalent to the other by a suitable
transformation in the jump probability distribution. We apply these
results to a recently proposed stochastic model describing the
dynamics of primary soil salinization, in which the salt mass
balance within the soil root zone requires the analysis of
different prescriptions arising from the resulting stochastic
differential equation forced by multiplicative white Poisson noise
whose features are tailored to the characters of the daily
precipitation. A method is finally suggested to infer the most
appropriate prescription from the data.
\end{abstract}

\pacs{Valid PACS appear here}
\keywords{It\^{o} and Stratonovich \sep Langevin like Equation
\sep
Master Equation  \sep Multiplicative Poisson Process}



\maketitle
\newpage
\section{Introduction}
Intense and concentrated state-dependent forcing events may often be
modeled as multiplicative random jumps, taking place according to an
underlying point process. Unlike the additive case, which counts a
relatively vast literature
\cite{Snyder1975,iturbe1999a,laio2001,Daly2006,Swain2008,azaele2010},
state-dependent jumps have been less investigated
\cite{VanDenBroeck1983,Sancho1987,Zygadlo1993,Pirrotta2007,Denisov2009}, and usually the state-dependency is assumed to be in the frequency of the jump occurrence, rather than in its amplitude.
The generalized Langevin equation (GLE) for white multiplicative
noise $\zeta(t)$, which can be either Gaussian or non
Gaussian,
\begin{equation}\label{sde}
   \dot{x}(t)\,=\,a(x,t)+\, b(x) \zeta(t),
\end{equation}
is ill-defined unless a prescription for the evaluation of the
stochastic term $b(x)\zeta(t)$ is specified \cite{Hanggi1982}. While
this issue is well understood for Gaussian white noise
(GWN) \cite{VanKampen1981}, a precise characterization of the noise
prescriptions and a clear connection between the different
interpretations are still missing for other kind of noises.

The last term in Eq. (\ref{sde}) for white Poisson (WP) process can
be written as $\zeta(t)= \xi_{\rho}(\nu,t)=\sum_i\,w_i\delta(t-t_i)$,
where the $t_i$ are the times at which jumps occur, $\delta$ is the
Dirac delta function, and the probability that $n$ jumps occur
during a time interval $\Delta t$ is given by the Poisson
distribution $P_n(t)=\exp{(-\nu \Delta t)}(\nu \Delta t)^n/n!$.
The jumps heights $w$ are independent and identically
distributed random variables with a probability distribution
function (PDF) $\rho(w)$. We note that the multiplicative
case $b(x) \zeta(t)$ of Eq. (\ref{sde}) is a special case, in which
the $x$ dependence of a more general state
dependent white noise $ \zeta(x,t)$ can be factorized out. Note that, while it is always
possible to reduce the state dependent noise as in (\ref{sde}) for
GWN, because GWN is fully characterized by its mean and variance,
this is not the case for the WP process.

The paper is organized as follows. First we show how different
prescriptions corresponding to the It\^{o} (I) and Stratonovich (S)
interpretation of a stochastic differential equation (SDE) arise naturally for
multiplicative jumps, depending on the relevant time scales of the
process. In section 3 we present the Master Equation (ME)
for a GLE with multiplicative compound Poisson process in both the
I and S prescriptions. The core of this
work is presented in section 4 where we show how in the linear case,
$b(x)\propto x$, the difference between prescriptions is properly
interpreted as a transformation of the jump size PDFs. We
demonstrate the relevance of these effects on a minimalist model of
soil salinization, describing possible long-term accumulation of
salt in soils in arid and semi-arid regions. In this problem,
the salt mass balance equation is characterized by state dependent
losses concentrated in negative jumps due to the leaching of salt
produced by intense rainfall events. The stochastic equation is
solved analytically obtaining explicitly the jump distributions that
arise in connection to the different noise interpretations.

\section{Connection between different prescriptions of a GLE and time scales of the process.}
\begin{figure}[htb]
  \includegraphics[width=29pc]{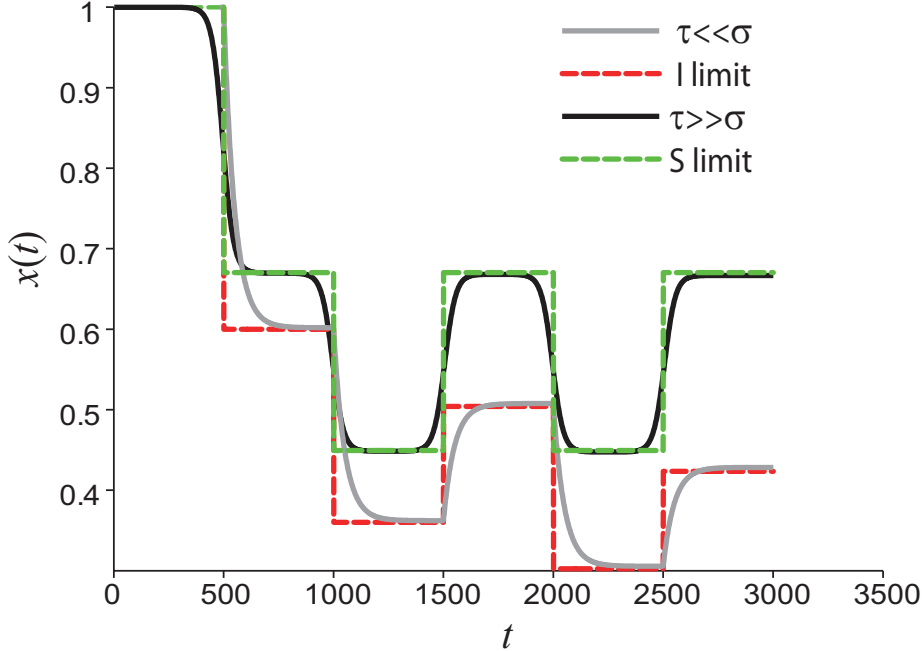}\\
  \caption{(Color online) Comparison between trajectories of a particle that undergoes impulsive multiplicative forcing in a viscosity field for different
  timescales ($\tau$ and $\sigma$), and the trajectories that result from the SDE $\dot{x}(t)=-x(t)\sum_{i=1}^{5}w_i\delta(t-t_i)$
  interpreted in the I and S prescriptions. The jumps in this case are given by $w_i=\pm 0.4$.}\label{fig1}
\end{figure}

We begin with a pedagogical example of a particle that
experiences multiplicative impulsive forcing events, proportional to
$\dot{\Theta}_{\tau}(t)$, of duration $\tau$, in a field
characterized by a friction coefficient $\psi$. Our
analysis is inspired by the work in references
\cite{Graham1982,Kupferman2004}. We choose
$\Theta_{\tau}(t)=\vartheta(t/\tau)$ with $\vartheta(z)\rightarrow 1 \,
(0)$ in the limit $z\rightarrow \infty \, (-\infty)$ so that
$\dot{\Theta}_{\tau}(t)\rightarrow \delta(t)$ in the $\tau
\rightarrow 0$ limit (in the distribution sense). We first consider
the case of a single jump event at $t=t_0>0$, where the dynamics is
described by the Newton equation
\begin{equation}\label{GLE_sj}
   m\,\ddot{x}(t)\,=\,-\psi\dot{x}+\,\psi b(x)w\dot{\Theta}_{\tau}(t-t_0),
\end{equation}
where the random jump $w$ is drawn from the jump size PDF $\rho(w)$.
Thus in Eq. (\ref{GLE_sj}) we have two time scales, $\sigma=m/\psi$
and $\tau$. The former is associated with the relaxation time toward
stationarity, while the latter is related to the characteristic
duration of the impulsive forcing. Different prescriptions of Eq. (\ref{sde}) arise depending on how the two emerging timescales
$\sigma$ and $\tau$ in Eq. (\ref{GLE_sj}) go to zero, i.e.
$\sigma\rightarrow0$ followed by $\tau\rightarrow0$ or viceversa (see Fig. \ref{fig1}).
For this reason, writing $\dot{x}(t)=b(x)w \delta(t-t_0)$ is
ambiguous, being the result of two different limit procedures with
different physical and mathematical meaning.

When $\sigma\ll\tau$ and then the zero limit of $\tau$ is taken in
Eq. (\ref{GLE_sj}), the S prescription of the SDE (\ref{sde}), which
preserves the usual rules of calculus, is obtained
\cite{stratonovich1963,vankampen2007}.
For example if $b(x)=x$, the resulting S-equation
$d\ln(x)/dt\,=\,w\dot{\Theta_{\tau}}(t-t_0)$, after performing the
limit $\tau\rightarrow0$, has formal solution
$x(t)=\left(1+\Theta(t-t_0)(e^{w}-1)\right)x_0$, where $x_0=x(0)$ and $\Theta$
is the Heaviside function. The corresponding PDF is
\begin{equation}\label{p(x,t)}
  p^{S}(x,t)=\delta(x-x_0)(1-\Theta(t-t_0))+\Theta(t-t_0)\frac{\rho\left(\ln(\frac{x}{x_0})\right)}{x}\Theta(\frac{x}{x_0}),
\end{equation} with initial condition $\rho^S(x,0)=\delta(x-x_0)$.
If otherwise $\tau\ll \sigma$, then Eq. (\ref{GLE_sj}) becomes
$\sigma\ddot{x}+\dot{x}=b(x)w\delta(t-t_0)$. Imposing the conditions
of continuity and right and left differentiability in $t_0$, the
initial conditions $x(t_0^-)$ and $\dot{x}(t_0^-)$, and taking the
limit $\sigma\rightarrow 0$, the solution is (again for the
case $b(x)=x$) $x(t)=x_0+x_{0}w\Theta(t-t_0)$. Note that the latter
corresponds to the solution in the It\^{o} prescription of the SDE
(\ref{GLE_sj}). From the formal It\^{o} solution of Eq. (\ref{GLE_sj}) we obtain the corresponding PDF in the I sense
\begin{equation}\label{P(x)}
p^{I}(x,t)=\delta(x-x_0)(1-\Theta(t-t_0))+\Theta(t-t_0)\frac{\rho\left(\frac{x-x_0}{x_0}\right)}{x_0}.
\end{equation}
The latter equation can not be made to correspond to Eq. (\ref{p(x,t)}) for
any choice of $\Theta(0)$. It is in fact interesting to observe that if we set
$\Theta(0)=\alpha$, then the parameter $\alpha$ defines where the
$b(x)$ that multiplies the jump is evaluated: when $\alpha=0$ $b(x)$ is
evaluated before the jump, while $\alpha=1/2$ corresponds to
calculating $b(x)$ in the middle of the jump. In the literature on GWN,
these choices are associated to the I and S prescriptions,
respectively \cite{gardiner2004,vankampen2007}. Conversely, as just
seen for a discrete jump process, the S interpretation of the SDE
(\ref{GLE_sj}) does not correspond to any of the $\alpha$
prescriptions. In other words, there is not an immediate intuitive
interpretation of the S prescription.

\section{Multiplicative Compound Poisson noise}
We generalize now our analysis to a process described by the following SDE,
\begin{equation}\label{GLE}
   \dot{x}(t)\,=\,a(x,t)\,+\, b(x) \xi^{\tau}_{\rho}(\nu,t),
\end{equation}
where
$\xi^{\tau}_{\rho}(\nu,t)=\sum_{i=1}^{N(t)}w\dot{\Theta}_{\tau}(t-t_i)$
is a colored compound Poisson processes (CP), with jump heights $w$,
each time drawn from a generic PDF $\rho(w)$, and $\{t_i\}$ are
random times whose sequence is drawn from a homogeneous Poisson
counting process $\{N(t),t\geq0\}$ of rate $\nu$. The case
in Section 2 corresponds to the special case of a finite deterministic number of jumps.
As before, the I interpretation consists of taking $\tau=0$ and, should
a jump occur at time $t$, of evaluating $b(x)$ at the r.h.s of Eq. (\ref{GLE}) before the jump occurrence, i.e. $x=x(t^-)$, while the
S interpretation of Eq. (\ref{GLE}) corresponds to performing the
zero limit of the correlation time $\tau$ of the colored Poisson
noise.

The S ME associated with the GLE (\ref{GLE}) can be derived through the generating function of $\xi^{\tau}_{\rho}(\nu,t)$ (see Appendix A), or in a more formal way
\citep{Hanggi1980,Sancho1987} as
\begin{equation}\label{ME_WP_strat}
  \frac{\partial P^S(x,t)}{\partial t}\,=\,\big[-\frac{\partial }{\partial x}a(x,t)
  +\nu\big\langle e^{-w\frac{\partial}{\partial x}b(x)}-1\big\rangle_{\rho(w)}\big]P^S(x,t),
\end{equation}
where $\langle\cdot\rangle$ denotes the ensemble average operator.
A simpler alternative derivation of the ME (\ref{ME_WP_strat}), can be obtained using the fact that in the S prescription the rules of calculus are preserved. Defining the function $\eta(x)\,=\,\int^x\frac{\mathrm{d}x'}{b(x')}$, the ME can be also written
as (see Appendix B):
\begin{equation}\label{ME_strat}
 \frac{\partial P^S(x,t)}{\partial t}\,=\,-\frac{\partial }{\partial x}\,\big[a(x,t)P^S(x,t)\big]\,
 +\,\nu\int_{-\infty}^{\infty}\frac{\rho(\eta(x)-\eta(x'))}{|b(x)|}P^S(x',t)dx'\,-\,\nu\, P^S(x,t).
\end{equation}

In the I prescription, $x(t)$ at time $t$ does not depend on
the noise $ \xi^{\tau=0}_{\rho}(\nu,t)\equiv\xi_{\rho}(\nu,t)$ at
the same time \cite{ito1951}. From this it follows that
\begin{equation}\label{ito_WP}
  \langle b(x) \xi_{\rho}(\nu,t)\rangle= \langle b(x)\rangle \langle \xi_{\rho}(\nu,t)\rangle.
\end{equation}
Therefore, if (\ref{GLE}) with $\tau=0$ is interpreted in the I sense, we can
change the size of the jumps from $w$ to $b(x)w$, and the corresponding
ME can be derived without ambiguity \cite{Hanggi1980,Denisov2009}
\begin{equation}\label{ME_ito}
 \frac{\partial P^I(x,t)}{\partial t}\,=\,-\frac{\partial }{\partial x}\,\big[a(x,t)P^I(x,t)\big]\,+\,\nu\int_{-\infty}^{\infty}\rho\left(\frac{x-x'}{b(x')}\right)\frac{P^I(x',t)}{|b(x')|}dx'\,-\,\nu\, P^I(x,t).
\end{equation}
Alternatively we achieved a different form of the I ME (\ref{ME_ito}) that is the I analogous of the S ME (\ref{ME_WP_strat})
 \begin{equation}\label{ME_ito_2}
  \frac{\partial P^I(x,t)}{\partial t}\,=\,-\frac{\partial }{\partial x}\,\big[a(x,t)P^I(x,t)\big]\,
  +\nu\big\langle :e^{-w\frac{\partial}{\partial x}b(x)}:-1\big\rangle_{\rho(w)}P^I(x,t),
\end{equation}
where $:\;:$ is an operator (analogous to the normal order operator in quantum field theory) which indicates that all the derivatives must be placed on the left of the expression, i.e. $:e^{-w\frac{\partial}{\partial x}b(x)}F(x):=\sum_{n=0}^{+\infty}\frac{(-w)^n}{n!}(\frac{\partial }{\partial x})^n(b(x)^nF(x))$. For details see Appendix C.

When $b(x)=b$ is constant, by using $ e^{-b\, w \frac{\partial }{\partial x}}P^S(x,t)=P^S(x-bw,t)$ the I and S MEs become coincident, as expected. In Appendix D we also show that, taking the limit $\nu\rightarrow\infty$, $\langle w \rangle\rightarrow0$, i.e. infinite frequency and infinitesimally small jumps, such that $\nu\langle w \rangle^2=D$ remains constant, Eqs. (\ref{ME_WP_strat}) and (\ref{ME_ito_2}) reduce to the well known I and S Fokker-Planck equation (FPE) for GWN  \cite{gardiner2004,vankampen2007}, respectively.

\section{Prescription-induced jump distributions}
It is clear from the previous MEs (\ref{ME_strat}) and
(\ref{ME_ito}) that the I and S prescriptions of the GLE
\begin{equation}\label{GLE1}
\dot{x}(t)=a(x,t)\,+\,b(x)\xi_{\rho}(\nu,t)
\end{equation} lead to different MEs.
We now want to determine the connection between the two different
interpretations. Specifically, we seek the two jump PDFs in the I
and S interpretation, $\rho_I$ and $\rho_S$, which give rise to the
same process. We also seek how to obtain one form when the other is
given. To this purpose it is sufficient to equate the two MEs,
(\ref{ME_strat}) and (\ref{ME_ito}) for simplicity, from which
\begin{equation}\label{IS_equiv2}
\frac{1}{|b(x')|}\rho_I(\frac{x-x'}{b(x')})\,=\,\frac{1}{|b(x)|}\rho_S(\eta(x)-\eta(x')).
\end{equation}

As a result, if Eq. (\ref{IS_equiv2}) can be
solved, given the jumps PDF and choosing the S (I) prescription for
Eq. (\ref{GLE1}), the solutions $\rho_I$ ($\rho_S$) of Eq. (\ref{IS_equiv2}) give the equivalent corresponding I (S) GLE and
ME. This is one of the main results of the paper and it
provides the connection between the prescription-induced jump
distributions $\rho_I$ and $\rho_S$, allowing link the It\^{o} ME
and the Stratonovich ME corresponding to a GLE with multiplicative
white Poisson noise.

The previous equation however has a solution only when $b(x)$ is a
linear function of $x$. To show this we rewrite Eq. (\ref{IS_equiv2}) as
\begin{equation}\label{IS_equiv2_r}
\rho_I(y)\,=\,\frac{|b(x')|}{|b(x)|}\rho_S(\eta(x)-\eta(x'))\equiv F(x',y),
\end{equation}
where $y=(x-x')/b(x')$. Because the l.h.s. of Eq. (\ref{IS_equiv2_r}) does not depend on $x'$, we must have
$\frac{\partial F}{\partial x'}=0$. If the latter condition holds
for all $\rho_S$, then we get $b''(x)=0$ whose solution is $b(x)=k x
+ c$ (see Appendix E for details). For other functional
shape of $b(x)$ the jumps PDF $\rho_I(w)$ depends also on the state
of the system, i.e., the dependence on $x$ of $\rho_I(w|x)$ cannot
be factored out. In this case, is not even clear to what a
Stratonovich prescription would correspond to.

Finally we derive the distribution of the impulses that may be
measured from the time series of the process (see inset in Fig.
\ref{fig2}). In fact, if a random jump (drawn from $\rho(w)$) occurs
at time $t$, then the size of the impulse that the whole process
experiences is $y_t=x(t+dt)-x(t)$. From the GLE (\ref{GLE}), we know
that with probability $\nu\,dt$, $\dot{x}= b(x) \,
w\,\Theta_{\tau}(t)$. Taking the limit $\tau\rightarrow 0$, and
using the definition of $\eta(x)$ we have (see Appendix B)
\begin{equation}\label{y(t)}
    y(t)\,=\,\left\{%
\begin{array}{ll}
    w \, b(x), & \hbox{(I)} \\
    \eta^{-1}(\eta(x)+w)-x & \hbox{(S)} \\
\end{array}%
\right.
\end{equation}and thus we obtain
\begin{equation}\label{ItoJ}
    \hat{P}^{I}(y,t)=\langle\delta(y-y_t)\rangle= \int_{-\infty}^{+\infty
 }\mathrm{d}x\mathrm{d}w\frac{1}{|b(x)|}P^I(x,t)\rho(w)\delta(w-y/b(x)) \\
\end{equation}\begin{equation}\label{StratJ}
 \hat{P}^{S}(y,t)= \int_{-\infty}^{+\infty}\mathrm{d}w\mathrm{d}xP^{S}(x,t)\rho(w)\frac{\delta(w-[\eta(x+y)-\eta(x)])}{|b(x+y)|},
\end{equation}i.e. the prescriptions characterize the PDF of the impulses of the whole process.

\section{Application to Soil Salinization}
The above mathematical problems naturally arise in the context of
the process of soil salinization. This is an extremely relevant
environmental problem as four million km$^2$ in arid and semi-arid
lands are affected by soil salinization, causing vegetation dieoff
and possible desertification \cite{Hillel2000,Suweis2009c}. In
natural salinization (unlike the anthropogenic one due to
irrigation), salt may accumulate in surface soils by dry and wet
deposition due to wind and rain. In this problem, state-dependent
Poisson jumps arise naturally when writing the salt mass balance
equation at the daily-to-monthly time scale for soil root zone used
as the control volume \cite{Suweis2009c}. Salt inputs due to
rainfall and wind act almost continuously in time, while the
state-dependent losses of salt occur through negative jumps due to
the leaching caused by intense rainfall events.
Schematically, the salt mass at time, $x(t)$, in the root
zone is described by the GLE:
\begin{equation}\label{salinitySDE}
    \frac{dx}{dt}\,=\,\Upsilon\,-\,x\,\xi_{\rho}(\nu,t),
\end{equation}
where $\Upsilon$ is the time-averaged salt mass input flux,
$\xi_{\rho}(\nu,t)$ is the leaching flux toward deeper layers, which
can be approximated by a WP process with $\rho(w)=\mu \exp(-\mu
w)\Theta(w)$. The leaching parameters $\nu$ (frequency of leaching
events) and $\mu$ (mean jump) can be expressed in terms of the
climatic, soil and vegetation properties \cite{Suweis2009c}. Because
the typical duration of leaching events is on the order of a few
hours, while the equilibration times of salt in the soil solution
(proportional to the inverse of its dissolution rate) tend to be
smaller (minutes to hours), this means that the inertia in the
dynamics is small ($\sigma\ll\tau$) and the physically correct
interpretation is likely to be the Stratonovich one.
\begin{figure}[ht]
 \includegraphics[width=29pc]{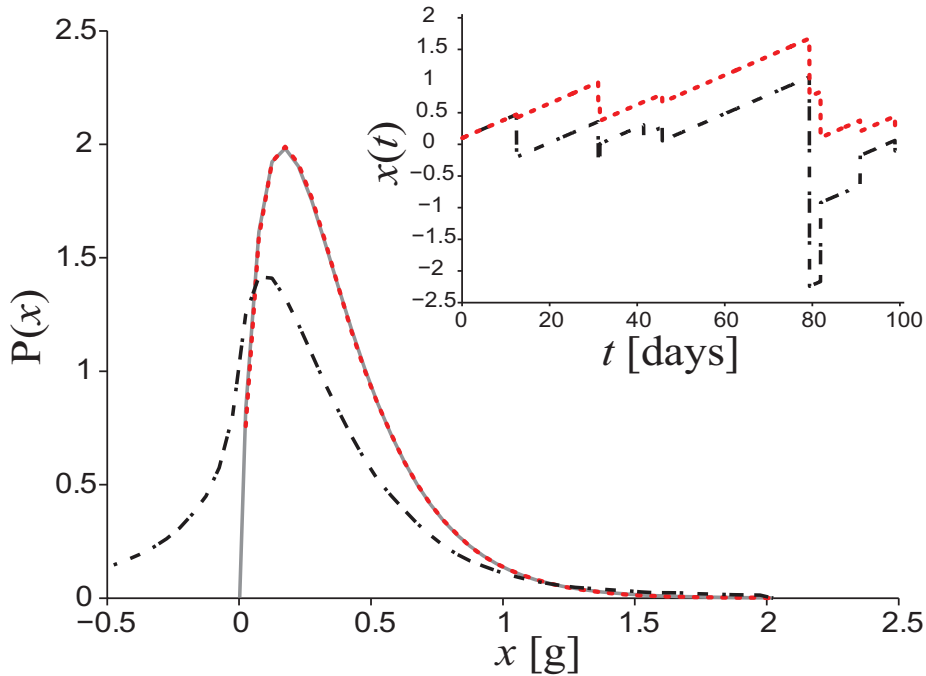}\\
  \caption{(Color online) Comparison of the steady state PDF of Eq. (\ref{salinitySDE}):  S solution (solid line, obtained analytically from Eq. (\ref{Strat_P})), I solution (dash-dot, from numerical simulation), I solution using the jump distribution given by Eq. (\ref{pres_iduc_dist1}) (dotted line, from numerical simulation). The numerical simulations confirm our analytical results. Inset: simulated trajectory of the salt mass under the two different prescriptions. Note that if artificial reflecting barriers are not imposed, the salt mass given by the I prescription of Eq. (\ref{salinitySDE}) may assume unphysical negative values. The parameters used for the simulation are $\mu=0.463$, $\nu=0.15\,day^{-1}$ and $\Upsilon=30\,mg/day$.}\label{fig2}
\end{figure}

The stationary solution of Eq. (\ref{ME_WP_strat}) in the S
prescription is a Gamma distribution (Fig.
\ref{fig2}) \cite{VanDenBroeck1983,Suweis2009c}
\begin{equation}\label{Strat_P}
P^S(x)\,=\,\mathcal{N}\,e^{-(x\, \nu\,/\,\Upsilon)}\,
x^{1\,/\,\mu},
\end{equation}
for $x>0$ and where $\mathcal{N}=(\frac{\nu}{\Upsilon})^{\frac{1+\mu}{\mu}}/\Gamma(\frac{1+\mu}{\mu})$
is the normalization constant and $\Gamma(z)$ the complete gamma function of argument $z$.
Eq. (\ref{Strat_P}) summarizes the soil salinity statistics as a
function of climate, soil and vegetation parameters, which may in
turn be used in conjunction with the soil moisture statistic to
obtain a full characterization of the salt concentration in the
root zone and the ensuing risk of salinization \cite{Suweis2009c}.

\begin{figure}
  \includegraphics[width=29pc]{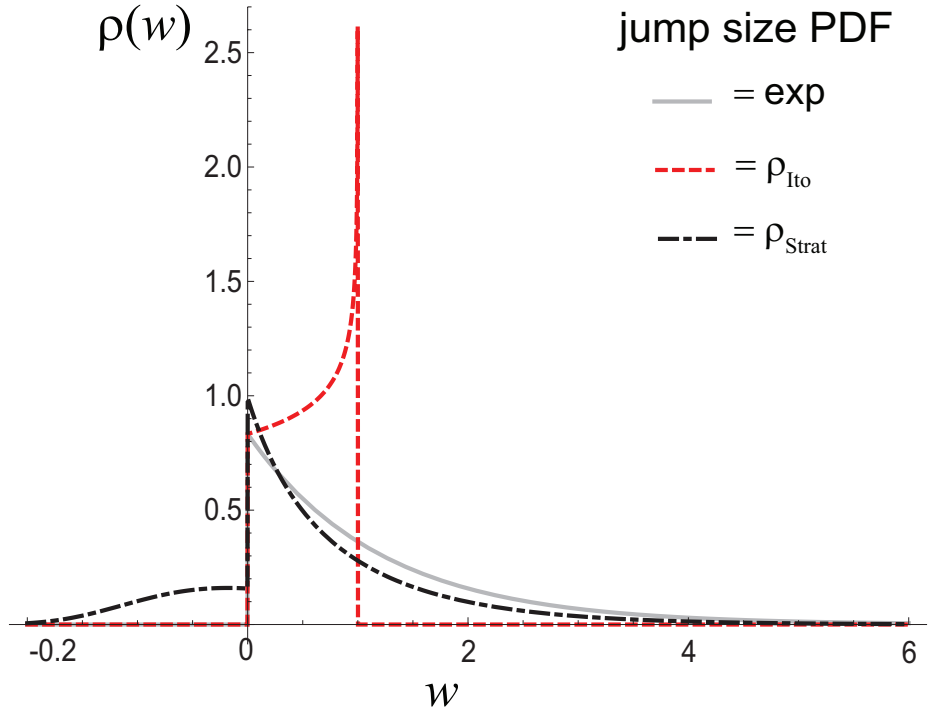}\\
  \caption{(Color online) Comparison between a jump exponential distribution $\rho(w)$ with mean $1/\gamma=0.8$, and the solutions $\rho_I(z)$, $\rho_S(z)$ of the prescription induced jumps corresponding to Eqs. (\ref{pres_iduc_dist1}) and (\ref{pres_iduc_dist2}) respectively, corresponding to the given $\rho(w)$.}\label{fig3}
\end{figure}

From Eq. (\ref{StratJ}) it is possible to derive the PDF of the impulses of the process for the S interpretation as
\begin{equation}\label{SalinityJumps}
    \hat{P}^{S}(y)\,=\, \epsilon e^{\epsilon y}\Theta(-y),
\end{equation}
which is an exponential distribution controlled by the parameter
$\epsilon=\nu/\Upsilon$, given by the ratio between the rate of
leaching events and the average rate of salt input. Thus if time
series of the process are available, the Stratonovich
assumption can be checked by backtracking information on the physical timescales
involved in the process, via a comparison with experimental
data. A further support for the S interpretation of Eq. (\ref{salinitySDE}) is given by the fact that $x$ must remain
positive after a jump, a fact that is not ensured by the I
interpretation unless a reflecting boundary in $x = 0$ is imposed
(see Fig. \ref{fig2}). We also computed the prescription-induced
jump distributions correspondence for this case ($b(x)=-x$), which
is $
  \rho_S(\ln|\frac{x'}{x}|)\,=\,|\frac{x}{x'}|\rho_I(1-\frac{x}{x'}),
$
where $x'$ and $x$ are the variables before and after the jump, respectively. By taking into account that in the S prescription $x,x'>0$, the I-jump PDF equivalent to $\rho_S(w)=\gamma e^{-\gamma w}\Theta(w)$ is
\begin{equation}\label{pres_iduc_dist1}
    \rho_I(z)=\gamma(1-z)^{\gamma-1}, \qquad z\in[0,1].
\end{equation}
This equivalence is indeed remarkable because it considerably facilitates the numerical simulation
of the salinity equation in the S formulation (see Fig. \ref{fig2}). On the other hand, if the GLE (\ref{salinitySDE}) were interpreted in the I sense, the ratio $x/x'$ could also be negative and the solution of Eq. (\ref{IS_equiv2}), for $\rho_I=\gamma\Theta(w)e^{-\gamma w}$, would read
\begin{equation}\label{pres_iduc_dist2}
     \rho_S(w)\,=\,\gamma e^{-\gamma -w}\left[\Theta(w)e^{\gamma e^{-w}} + e^{-\gamma e^{-w}}\right] \ \ \ w\in]-\infty,+\infty[.
\end{equation}This implies that possible negative jumps (that occur for $x<0$) in the I prescription for the given $\rho_I(w)$, would be explicitly present in the corresponding equivalent S-jump PDF $\rho_S(w)$ (see Fig. (\ref{fig3})).\\

\section{Conclusions}

In this work we have proposed a novel approach to solve the
Ito-Stratonovich (I-S) dilemma for GLE with multiplicative WP noise.
We have shown how different interpretations lead to different
results and that choosing between the I and S prescriptions is
crucial to describe correctly the dynamics of the model systems, and
how this choice can be determined by physical information about the
timescales involved in the process. Moreover, we have addressed the
related issue of finding a connection between the I and S
interpretations in the case of linear WP noise. Differently from the
introduction of a drift previously proposed
\cite{Zygadlo1993,Pirrotta2007}, we have found such connection in a
transformation of the jumps PDFs and tested our results
numerically. Our results are also consistent with the physics of the
random forcing, which takes place at specific points in time,
whereas a continuously-acting spurious drift would conceptually
violate the causality of the process. In particular, once the GLE
(\ref{GLE1}) is given, its I and S interpretations are shown to be
equivalent if $\rho_I$ and $\rho_S$ satisfy the prescription-induced
jumps PDF Eq. (\ref{IS_equiv2}). The case of nonlinear
multiplicative WP noise will be studied elsewhere. We have applied
our results to the geophysical problem of soil salinization, by
solving a minimalist model that describes the salt mass and
concentration in a soil control volume as a function of climatic and
ecohydrological parameters.

\begin{acknowledgments}
This research is supported by funds provided by the ERC Advanced
Grant RINEC-227612 and SFN funding through project 200021-124930/1.
AP acknowledges the Landolt Visiting program at EPFL for financial
support. AM acknowledges funds provided by Fondazione Cariparo
(Padova, IT).
\end{acknowledgments}
\newpage

\section{Appendix A}
 The stochastic process under study is described by the GLE (\ref{GLE}) presented in the main text.
For simplicity in the following we have set $\xi_{\rho}^{\tau}(t,\nu)=\xi(t)$.
The CP is characterized by the correlation structure ( $\langle\cdot\rangle$ denotes the ensemble average)
\begin{equation}\label{covCP}
    \langle\xi(t)\xi(t+\tau)\rangle \,\sim \,  e^{-\frac{t}{\tau}},
\end{equation}where $\tau$ is the characteristic time of the process and we have omitted all the sub- and super- scripts to simplify the notation. If
$\Phi_t$ is the generating function of CP at time $t$, then
\begin{eqnarray}
  \Phi_t[v] &=& \bigg\langle e^{i\int_0^t v(s)\xi(s)ds}\bigg\rangle\,=\,e^{\Psi_t[v]} \\
  \nonumber &=& \sum_{n=0}^{+\infty}e^{-\nu t}\frac{(\nu t)^n}{n!}\int dw\rho(w)\int\prod_{j=1}^{n}\frac{dt_j}{t}e^{i\sum_{j=1}^n w_j\int_0^t v(s)\Theta_{\tau}(s-t_j)ds} \\
  \nonumber &=&  \sum_{n=0}^{+\infty}e^{-\nu t}\frac{\nu^n}{n!}\bigg[\int_0^t dr\int dw \rho(w)\exp\bigg[i\sum_{j=1}^{n}w_j\int_0^t
  v(s)\Theta_{\tau}(s-t_j)ds\bigg]\bigg].
\end{eqnarray}
Moreover if we define $\hat{\rho}=\int_{-\infty}^{+\infty} e^{i v w}\rho(w)dw$ as the characteristic function of
$\rho(w)$, then we have
\begin{equation}\label{gf}
\Phi_t[v]\,=\,\exp\bigg[-\nu t + \nu\int_0^t dr
\hat{\rho}\left(\int_0^t v(s)\Theta_{\tau}(s-\tau)d\tau\right)\bigg],
\end{equation}and thus
\begin{equation}\label{log_gf}
\Psi_t[v]\,=\,\ln\Phi_t[v]\,=\,\nu\int_0^t dr\bigg[
\hat{\rho}\left(\int_0^t v(s)\Theta_{\tau}(s-\tau)d\tau\right)-1\bigg].
\end{equation}
The Stratonovich interpretation of Eq. (\ref{GLE1}) arises when the
limit $\Theta_{\tau}(t-\tau)\rightarrow\delta(t-\tau)$ is taken
\cite{stratonovich1963,VanKampen1981}, that is considering a white
Poisson process (WP) as the zero limit of the correlation time of
the corresponding CP. For a WP the logarithm of the generating
function thus reads
\begin{equation}\label{log_gf_WP}
\Psi_t[v]\,=\,\nu\int_0^t dr\big[\hat{\rho}(v(r))-1\big].
\end{equation}Finally because of the Kubo theorem \cite{Kubo1962}
\begin{equation}\label{Kubo_th}
\Psi_t[v]\,=\,\sum_{n=1}^{\infty}\frac{i^n}{n!}\int_0^t ds_1\cdots
ds_n v(s_1)\cdots
v(s_n)\langle\langle\xi(t_1)\cdots\xi(t_n)\rangle\rangle_n,
\end{equation}where $\langle\langle\cdots\rangle\rangle_j$ is the j-th cumulant,
i.e., $\langle\langle\cdot\rangle\rangle_1=\langle\cdot\rangle$,
$\langle\langle\cdot\cdot\rangle\rangle_2=\langle\cdot\cdot\rangle-\langle\cdot\rangle\langle\cdot\rangle$,
etc...

From Eqs. (\ref{log_gf_WP}) and (\ref{Kubo_th}) we obtain the
explicit formula to calculate the cumulants
\begin{eqnarray}
\label{cum1}  \langle\langle\xi(t_1)\rangle\rangle_1 &=& \frac{\delta\Psi_t}{i\delta v(t_1)}\,=\,\frac{\nu}{i}\hat{\rho}'(v)_{|_{v=0}} \\
\label{cum2}  \langle\langle\xi(t_1)\xi(t_2)\rangle\rangle_2 &=& \frac{\delta^2\Psi_t}{i^2\delta v(t_1)\delta v(t_2)}\,=\,\frac{\nu}{i^2}\hat{\rho}''(v)_{|_{v=0}}\delta(t_2-t_1) \\
\label{cumn}
\langle\langle\xi(t_1)\cdots\xi(t_n)\rangle\rangle_n
&=&\frac{\delta^n\Psi_t}{i^n\delta v(t_1)\cdots\delta
  v(t_n)}\,=\,\frac{\nu}{i^n}\hat{\rho}^{(n)}(v)_{|_{v=0}}\delta(t_2-t_1)\cdot\cdots\delta(t_n-t_{n-1}).
\end{eqnarray}
In this way, once $\rho(w)$ is given, we have a complete description
of the WP. For example in the case of exponential distributed jumps,
i.e. $\rho(w)=\frac{1}{\langle w\rangle} e^{-\frac{w}{\langle
w\rangle}}$, the WP is fully characterized by the moments
\begin{eqnarray}
 \langle \langle\xi(t)\rangle\rangle_1 &=& \nu\langle w\rangle \\
  \langle\langle\xi(t_1)\xi(t_2)\rangle\rangle_2 &=& \nu\langle w^2\rangle\delta(t_1-t_2) \\
  \langle\langle\xi(t_1)...\xi(t_n)\rangle\rangle_n &=&  \nu\langle w^n\rangle\delta(t_1-t_2)...\delta(t_{n-1}-t_n).
\end{eqnarray}

Once we have calculate all the moments of the WP process we can
easily achieve the ME corresponding to the GLE
(\ref{GLE1}). For a given realization of $\xi$ the
solution of Eq. (\ref{GLE}) is
\begin{equation}\label{part_sol}
p^S(x,t|\xi)=\delta(x-x(t)).
\end{equation}To obtain the general solution of Eq. (\ref{GLE1}) we simply have to take the ensemble average of different trajectories
\begin{equation}\label{gen_sol}
\langle p^S(x,t|\xi)\rangle = P^S(x,t).
\end{equation}Differentiating both sides of Eq. (\ref{part_sol}) and using Eq. (\ref{GLE1}) we have
\begin{eqnarray}
  \partial_t p^S(x,t|\xi)&=& \partial_x \delta(x-x(t))[-\dot{x}(t)] \\
   &=& -\partial_x \delta(x-x(t))[a(x(t),t) \,+ \,b(x(t)) \xi(t)] \\
   &=&  -\partial_x \delta(x-x(t))[a(x(t),t) \,+ \,b(x(t)) \xi(t)],
\end{eqnarray}
and thus we obtain a forward ME for the PDF conditioned by a given realization of the WP
\begin{equation}\label{op_ME}
 \frac{\partial}{\partial t} p^S(x,t|\xi)\,=\,-\mathcal{O}(x,\partial_x,t)\,p^S(x,t|\xi),
\end{equation}where $\mathcal{O}(x,\partial_x,t)=\partial_x [a(x(t),t) +b(x(t)) \xi(t)]$ is the forward time evolution operator.
The solution of Eq. (\ref{op_ME}), for the initial condition $ p^S(x(0),0|\xi)=\delta(x-x(0))$  is
\begin{equation}\label{sol_op_ME}
  p^S(x,t|\xi)\,=\, \mathrm{T}\left(\exp\big[-\int_{0}^t(\partial_{x}a(x,\tau)+\partial_{x}b(x(\tau))\xi(\tau))d\tau\big]\right)\delta(x-x(0)),
\end{equation}
where $\mathrm{T}$ is the T-product operator. Using Eq. (\ref{gen_sol}) and the Kubo relation (\ref{Kubo_th}), an explicit
formula for the general formal solution of the GLE (\ref{GLE1}) in
the Stratonovich prescription is obtained
\begin{eqnarray}\label{gen_sol_op_ME}
   P^S(x,t) &=&\mathrm{T}\bigg(\exp\big[-\int_{0}^t \partial_{x}a(x,\tau)d\tau - \sum_{n=1}^{\infty}\int_{0}^t dt_1\cdots\int_{0}^t dt_n\partial_{x}b(x(t_1))\cdots\partial_x b(x(t_n))\times\qquad\\
  &\times & \langle\langle\xi(t_1)\cdots\xi(t_n)\rangle\rangle\big]\bigg)\delta(x-x(0)).
\end{eqnarray}
Thanks to Eqs. (\ref{cum1}), (\ref{cum2}) and (\ref{cumn}) we have a
complete characterization of the cumulants, and thus substituting
Eq. (\ref{cumn}) into Eq. (\ref{gen_sol_op_ME}) we obtain
\begin{eqnarray}
  \nonumber P^S(x,t)&=& \mathrm{T}\left(\exp\big[-\int_{0}^t\partial_{x}a(x,\tau)d\tau+\sum_{n=1}^{\infty}\nu\int_{0}^t\left(-\partial_{x}b(x(\tau))\right)^n\hat{\rho(0)}^{(n)}d\tau\big]\right)\\
   \label{formal_sol_GLE} &=& \mathrm{T}\left(\exp\big[-\int_{0}^t\partial_{x}a(x,\tau)d\tau-\nu\int_{0}^td\tau\big\langle e^{-\partial_{x}b(x(\tau))}-1\big\rangle_{\rho(w)}\big]\right)
\end{eqnarray}
Eventually, differentiating Eq. (\ref{formal_sol_GLE}) with respect
to $t$ we obtain the ME corresponding to the GLE (\ref{GLE1}) in the
Stratonovich interpretation
\begin{equation}\label{ME_WP_strat_A}
  \frac{\partial P^S(x,t)}{\partial t}\,=\,\bigg[-\frac{\partial }{\partial x}a(x,t)+\nu\big\langle e^{-w\frac{\partial}{\partial x}b(x)}-1\big\rangle_{\rho(w)}\bigg]P^S(x,t),
\end{equation}that is the ME (\ref{ME_WP_strat}) reported in the main text.

\section{Appendix B}
We now show the derivation of the S ME (\ref{ME_strat}) in the main text and its equivalence with Eq. (\ref{ME_WP_strat}).
We can write the GLE (\ref{GLE}) as
\begin{equation}\label{single_jump_Poisson_Strat}
 \dot{x}(t) = \left\{
  \begin{array}{ll}
    a(x,t), & \hbox{with probability $1\,-\,\nu\,dt$;} \\
    b(x) \, w\,h_{\tau}(t) , & \hbox{with probability $\nu\,dt$,}
  \end{array}
\right.
\end{equation}where $h_{\tau}(t)\,=\,\dot{\Theta}_{\tau}(t)$. We now consider only the effect of the jumps on $x$.
From Eq. (\ref{single_jump_Poisson_Strat}) we have that $dx/b(x(t))\,=\,w\,h_{\tau}(t)dt$, and setting
\begin{equation}\label{mu(x)}
    \frac{d\eta(x)}{dx}\,=\,\frac{1}{b(x)}\quad\Rightarrow\quad \eta(x)\,=\,\int^x\frac{dx'}{b(x')},
\end{equation}Eq. (\ref{single_jump_Poisson_Strat}) becomes
\begin{equation}\label{GLE_mu}
    d\eta(x(t))\,=\,w\,h_{\tau}dt,
\end{equation}which integrated between $t$ and $t+dt$ reads
\begin{equation}\label{GLE_mu_int}
    \eta(x(t+dt))\,=\,\eta(x(t))\,+\,w\,\Delta\Theta_{\tau}(t)\quad\Rightarrow\quad x(t+dt)\,=\,\eta^{-1}\big[\eta(x(t))\,+\,w\,\Delta\Theta_{\tau}(t)\big],
\end{equation}where $\Delta\Theta_{\tau}(t)=\Theta_{\tau}(t+dt)-\Theta_{\tau}(t)$.\\
Finally we can write the discrete ME corresponding to the GLE (\ref{GLE1}) interpreted in the Stratonovich sense
\begin{eqnarray}\label{discr_ME_strat}
    P^S(x,t+dt)&=&(1-\nu)\, dt\,\int_0^{\infty}dx'\,P^S(x',t)\,\delta\left(x-(a(x')dt+x')\right)\,+\\\nonumber
    & &+\,\nu\, dt \int_0^{\infty}\int_0^{\infty}\rho(w)P^S(x',t)\delta\left(x-(\eta^{-1}[\eta(x')\,+\,w])\right)\,dw\,dx',
\end{eqnarray}
where we have performed the limit $\tau\rightarrow0$ of the GLE (\ref{GLE}) and used the fact that $\lim_{\tau\rightarrow0}\Delta\Theta_{\tau}(t)=1$.
The integral in the r.h.s of Eq. (\ref{discr_ME_strat}) can be rewritten, inverting the Dirac Delta with respect to $w$ and using the rule of the inverse function, as $\int_0^{\infty}\int_0^{x}\rho(w)P^S(x')\frac{\delta\left(w-(\eta(x)-\eta(x'))\right)}{|1/\eta'(x)|}\,dw\,dx'$ and thus, after taking the continuum time limit, the Master Equation (\ref{discr_ME_strat}) becomes
\begin{equation}\label{ME_strat_A}
 \frac{\partial P^S(x,t)}{\partial t}\,=\,-\frac{\partial }{\partial x}\,\big[a(x,t)P^S(x,t)\big]\,+\,\nu\int_0^{\infty}\frac{\rho(\eta(x)-\eta(x'))}{|b(x)|}P^S(x',t)dx'\,-\,\nu\, P^S(x,t),
\end{equation}that is Eq. (\ref{ME_strat}) reported in the main text.

In order to show the equivalence between Eqs. (\ref{ME_WP_strat}) and (\ref{ME_strat}) we define
\begin{equation}\label{Q}
 Q(x,w)\,=\,\int_0^{\infty}P^S(x')\delta\left(x-(\eta^{-1}[\eta(x')\,+\,w])\right)\,dx',
\end{equation}
so we have that the integral in Eq. (\ref{ME_strat_A}) is simply $\int_0^{\infty} Q(x,w)\rho(w)dw$.

Differentiating Eq. (\ref{Q}) with respect to $w$ we obtain
the partial differential equation
\begin{equation}\label{pde_Q}
\partial_w\,Q(x,w)\,=\,-\partial_x\,b(x)Q(x,w)=-\mathcal{H}Q,
\end{equation}
where we used Eqs. (\ref{mu(x)}) and (\ref{Q}) and the definition of the derivative of the inverse function.
The solution of Eq. (\ref{pde_Q}) is
\begin{equation}\label{pde_Q_sol}
Q(x,w)\,=\,e^{-w\mathcal{H}}Q(x,0)=e^{-w\partial_x b(x)}P^S(x).
\end{equation}
We thus have
\begin{equation}\label{ME_eq}
\int_0^{\infty}\frac{\rho(\eta(x)-\eta(x'))}{|b(x)|}P^S(x',t)dx'=\big\langle e^{-w\partial_x b(x)}\big\rangle_{\rho}P^S(x,t),
\end{equation}
which substituted in Eq. (\ref{ME_strat_A}) proves the equivalence between the MEs (\ref{ME_WP_strat}) and (\ref{ME_strat}) in the main text.
\section{Appendix C}
We present in this Appendix the derivation for the I ME (\ref{ME_ito_2}) and its equivalence with Eq. (\ref{ME_ito}).
We first note that the integral in the ME (\ref{ME_ito}) can be rewritten as
\begin{equation}\label{ito_ME_int}
    \int_{-\infty}^{\infty}\rho\left(\frac{x-x'}{b(x')}\right)\frac{P^I(x',t)}{|b(x')|}dx'=\int_{-\infty}^{\infty}\int_{-\infty}^{\infty}\rho(w)\delta(x-x'-wb(x'))P^I(x',t)dx'dw
\end{equation}
Formally expanding the Dirac delta
\begin{equation}\label{dirac_exp}
\delta(x-x'-wb(x'))=\sum_{n=0}^{+\infty}\frac{(-w)^n}{n!}(\frac{\partial}{\partial x})^nb(x')^n\delta(x-x')
\end{equation}
and substituting Eq. (\ref{dirac_exp}) in (\ref{ito_ME_int}) we have
\begin{equation}\label{ito_ME_int2}
    \int_{-\infty}^{\infty}\rho\left(\frac{x-x'}{b(x')}\right)\frac{P^I(x',t)}{|b(x')|}dx'=\big\langle(:e^{-w\frac{\partial}{\partial x}b(x)}:)P^I(x,t)\big\rangle_{\rho},
\end{equation}
where (:$e^{-w\frac{\partial}{\partial x}b(x)}$:)$P^I(x,t)\equiv\sum_{n=0}^{+\infty}\frac{(-w)^n}{n!}(\frac{\partial}{\partial x})^nb(x)^n P^I(x,t)$. Using the expression (\ref{ito_ME_int2}) in Eq. (\ref{ME_ito}) we obtain the ME (\ref{ME_ito_2}).

\section{Appendix D}
We now derive the well known FPE corresponding to the GLE (\ref{GLE}) when $\xi(t)$ is a GWN with mean $\langle \xi(t) \rangle=0$ and correlation $\langle \xi(t)\,\xi(s) \rangle=2D\,\delta(t-s)$, from the MEs (\ref{ME_WP_strat}) and (\ref{ME_ito_2}) presented in the main text. We generalize our results to any jump size PDF of the form
\begin{equation}\label{rhotof}
    \rho(w)=\gamma f(\gamma w),
\end{equation}
with $\gamma>0$ and $\int w^n \rho(w)dw=\langle w^n \rangle_{\rho}<\infty$ $\forall n$. We note that the latter condition implies $\gamma\int dw w^n f(\gamma w)=\gamma^{-n}\int dz z^n f(z)=\gamma^{-n}\langle z^n \rangle_{f}<\infty$ $\forall n$.\\
\emph{Stratonovich Eq.}
The case for the Stratonovich prescription has been first presented in \cite{VanDenBroeck1983}.
The FPE corresponding to multiplicative GWN process interpreted in the Stratonovich sense is
\begin{equation}\label{FP_s}
    \frac{\partial}{\partial t}P^S(x,t)\,=\,-\frac{\partial}{\partial x}\big[a(x,t)\,P^S(x,t)\big]\,+\,D\,\frac{\partial}{\partial x}b(x)\frac{\partial}{\partial x}b(x)\,P^S(x,t).
\end{equation}
Once we consider a zero mean WP process, the ME (\ref{ME_WP_strat}) reads as \cite{VanDenBroeck1983}
\begin{eqnarray}\label{ME_strat_Gaus0}
\frac{\partial P^S(x,t)}{\partial t}&=&-\frac{\partial }{\partial
x}\bigg[\big[a(x,t)-\nu\langle w \rangle b(x)\big]P^S(x,t)\bigg]\,+\,\nu\big\langle e^{-w\frac{\partial}{\partial x}b(x)}-1\big\rangle_{\rho} P^S(x,t)\\
\label{ME_strat_Gaus1}&=&-\frac{\partial }{\partial x}\big[a(x,t)P^S(x,t)\big]\,+\,\nu \sum_{n=1}^{+\infty}(-\frac{1}{\gamma})^n\frac{\langle z^n \rangle_f}{n!}\big(\frac{\partial}{\partial x}b(x)\big)^n P^S(x)
\end{eqnarray}
where the integral in the r.h.s of Eq. (\ref{ME_strat_Gaus0}) has been expanded as
\begin{equation}\label{ME_strat_int}
\big\langle e^{-w\frac{\partial}{\partial x}b(x)}\big\rangle_{\rho} P^S(x)=\sum_{n=0}^{+\infty}(-1)^n\frac{\langle w^n \rangle}{n!}\big(\frac{\partial}{\partial x}b(x)\big)^n P^S(x)=\sum_{n=0}^{+\infty}(-\frac{1}{\gamma})^n\frac{\langle z^n \rangle_f}{n!}\big(\frac{\partial}{\partial x}b(x)\big)^n P^S(x).
\end{equation}
Taking the limit  $\nu,\gamma\rightarrow\infty$, such that $\frac{\nu}{\gamma^2}=D'$, then $\frac{\nu}{\gamma^n}\rightarrow0$ for $n>2$ and the latter ME (\ref{ME_strat_Gaus1}) corresponds exactly to the FPE (\ref{FP_s}) with $D=D'\frac{\langle z \rangle_f}{2}$.\\
\emph{It\^{o} Eq.}
The FPE corresponding to multiplicative GWN process interpreted with the It\^{o} prescription is
  \begin{equation}\label{FP_i}
    \frac{\partial}{\partial t}P^I(x,t)\,=\,-\frac{\partial}{\partial x}\big[a(x,t)\,P(x,t)\big]\,+\,D\,\frac{\partial^2}{\partial x^2}\big[b(x)^2\,P^I(x,t)\big].
\end{equation}
We now repeat the same procedure as before, starting from the zero mean I ME
\begin{equation}\label{ME_ito_A}
 \frac{\partial P^I(x,t)}{\partial t}\,=\,-\frac{\partial }{\partial x}\,\big[[a(x,t)-\nu \langle w \rangle b(x)]P^I(x,t)\big]\,+\,\nu\big\langle (:e^{-w\frac{\partial}{\partial x}b(x)}:)-1\big\rangle_{\rho} P^I(x,t).
\end{equation}
We can expand the r.h.s. remembering that the operator $::$ means that all the derivatives must be placed on the left of the expression
\begin{equation}\label{int_ito}
 \nu \big\langle (:e^{-w\frac{\partial}{\partial x}b(x)}:)-1\big\rangle_{\rho} P^I(x,t)=-\nu \langle w \rangle_{\rho} \frac{\partial}{\partial x}[b(x)P^I(x,t)]+\nu \sum_{n=2}^{+\infty}(-\frac{1}{\gamma})^n\frac{\langle z^n \rangle_f}{n!}\big(\frac{\partial}{\partial x}\big)^n[b(x)^n P^I(x,t)]
\end{equation}
Eventually, inserting Eq. (\ref{int_ito}) in the I ME (\ref{ME_ito_A}) and taking
$\nu,\gamma\rightarrow\infty$ with
$\frac{\nu}{\gamma^2}=D'$ and  $D=D'\frac{\langle z \rangle_f}{2}$ we obtain the I FPE (\ref{FP_i}).

\section*{Appendix E}
In this section we show how a solution of Eq. (\ref{IS_equiv2}), rewritten as
\begin{equation}\label{IS_equiv2_r2A}
\rho_I(y)\,=\,\frac{|b(x')|}{|b(x)|}\rho_S(\eta(x)-\eta(x'))\equiv F(x',y),
\end{equation}
where $y=(x-x')/b(x')$, exist only if is $b$ is a linear function.

Because the l.h.s. of Eq. (\ref{IS_equiv2_r2A}) does not depend on $x'$, we must have
$\frac{\partial F}{\partial x'}=0$, that explicitly read as
\begin{eqnarray}\label{dFdxequal01}\nonumber
0&=&\rho_S(\eta(x)-\eta(x'))\big[sgn[b(x')]\frac{b'(x')}{|b(x)|}-sgn[b(x)]\frac{b'(x)}{|b(x)|^2}|b(x')|(1+yb'(x'))\big]+\\
&+&\rho_S'(\eta(x)-\eta(x'))\frac{|b(x')|}{|b(x)|}\big[\eta'(x)(1+yb'(x)-\eta'(x'))\big].
\end{eqnarray}
The latter, using Eq. (\ref{mu(x)}), can be expressed as
\begin{equation}\label{dFdxequal02}
\frac{\rho_S'(\eta(x)-\eta(x'))}{\rho_S(\eta(x)-\eta(x'))}\big(\frac{1}{b(x)}(1+yb'(x'))
-\frac{1}{b(x')})+\frac{b'(x')}{b(x')}-\frac{b'(x)}{b(x)}(1+yb'(x'))=0.
\end{equation}

Eq. (\ref{dFdxequal02}) must hold for all $\rho_S$, then the solution of Eq. (\ref{dFdxequal02}) is given by the function $b$ that satisfies the conditions
\begin{eqnarray}\label{dFdxequal03a}
b(x')(1+yb'(x'))&=&b(x)\\
\label{dFdxequal03b} b(x')b'(x)(1+yb'(x'))&=&b(x)b'(x').
\end{eqnarray}

Combining Eqs. (\ref{dFdxequal03a}) and (\ref{dFdxequal03b}) and using $x=b(x')y+x'$ we obtain the equation $b'(b(x')y+x')=b'(x')$. If we take the derivative of both side with respect to the independent variable $y$, then we have $b''(b(x')y+x')b(x')=0$. This implies $b''(x)=0$ $\forall x$, which solution is $b(x)=k x$ (with $k$ any constant).

\end{document}